\documentclass[fleqn,10pt]{wlscirep}
\usepackage[utf8]{inputenc}
\usepackage[T1]{fontenc}
\usepackage{siunitx}
\usepackage{multirow}
\usepackage{subcaption}
\usepackage{booktabs}
\usepackage{graphicx} 
\usepackage{float}

\usepackage{numprint}
\npdecimalsign{\ensuremath{.}}
\newcommand{\np}[1]{\numprint{#1}}

\title{Scientific Reports Title to see here}

\title{
The anatomy of Green AI technologies: structure, evolution, and impact}

\author[1,2]{Lorenzo Emer}
\author[1,3]{Andrea Mina}
\author[1,4,*]{Andrea Vandin}
\affil[1]{Institute of Economics and L'EMbeDS, Scuola Superiore Sant’Anna, Piazza Martiri della Libertà, 33, 56127 Pisa (Italy)}
\affil[2]{Department of Computer Science, University of Pisa, Largo B. Pontecorvo 3, 56126 Pisa (Italy)}
\affil[3]{Centre for Business Research, University of Cambridge, 11-12 Trumpington Street, CB2 1QA Cambridge (UK)}
\affil[4]{DTU Technical University of Denmark, Anker Engelunds Vej 101, 2800, Kongens Lyngby (Denmark)}

\affil[*]{Corresponding author: Andrea Vandin, andrea.vandin@santannapisa.it}
\usepackage{xcolor}

\begin{abstract}

Artificial intelligence (AI) is a key enabler of innovation against climate change. In this study, we investigate the intersection of AI and climate adaptation and mitigation technologies through patent analyses of a novel dataset of approximately \np{63000} “Green AI” patents. 
We analyze patenting trends, corporate ownership of the technology, the geographical distributions of patents, their impact on follow-on inventions and their market value. We use topic modeling (BERTopic) to identify 16 major technological domains, track their evolution over time, and identify their relative impact. We uncover a clear shift from legacy domains such as combustion engines technology to emerging areas like data processing, microgrids, and agricultural water management. We find evidence of growing concentration in corporate patenting against a rapidly increasing number of patenting firms. Looking at the technological and economic impact of patents, while some Green AI domains combine technological impact and market value, others reflect 
weaker private incentives for innovation, despite their relevance for climate adaptation and mitigation strategies. This is where policy intervention might be required to foster the generation and use of new Green AI applications.  
\end{abstract}
\begin{document}

\flushbottom
\maketitle
%
%
\thispagestyle{empty}


\section*{Introduction: What is Green AI, and why does it matter?}

\newcolumntype{H}{>{\setbox0=\hbox\bgroup}c<{\egroup}@{}}

Climate change is one of the defining challenges of the 21st century. Between 2011 and 2020, the average global temperature rose by 1.1°C above pre-industrial levels, while atmospheric CO\textsubscript{2} levels reached the highest concentrations in over two million years \cite{IPCC2023}. Exceeding the 2°C threshold risks triggering irreversible damage to ecosystems, public health, and economic systems, while exacerbating global inequalities \cite{Maule2017, Smith2022}. Addressing this crisis demands not only political will but also technological innovation, particularly in transforming energy systems, industrial processes, and climate resilience infrastructures \cite{Rogelj2015}.

Artificial intelligence (AI) has emerged as a potentially transformative tool in this context. Defined as the ability of systems to learn from data and adapt to achieve specific goals \cite{kaplan2019siri}, AI is now applied across a range of climate-related areas of technological research. These include forecasting energy production from renewable sources, optimizing industrial efficiency, improving disaster response, assisting water and agricultural management, and supporting carbon capture and geoengineering technologies \cite{Cowls2023}. Despite the breadth of potential applications, the empirical study of “Green AI” (that is, AI-driven technologies for climate mitigation and adaptation) remains limited and scattered across sectors.

This research contributes to filling this gap in the literature through the analysis of patent data. Patents offer a valuable lens on technological development, combining structured information across countries, sectors, and time periods \cite{hall2001nber, Benson2015}. We construct a novel dataset of approximately \num{63000} U.S. patents, spanning from 1976 to 2023, that are both classified as artificial intelligence inventions and as climate-related technologies, using the Cooperative Patent Classification system (Y02/Y04S) and recent AI patent identification methods, namely the Artificial Intelligence Patent Dataset (AIPD)\cite{pairolero2024aipd}. These patents, which represent the intersection of AI and environmental technologies, constitute a rapidly growing yet underexplored class of innovations.

A key contribution of this paper is to uncover the main technological domains within the universe of US Green AI patents by applying topic modeling. This technique is ideally suited to this task as an unsupervised Natural Language Processing (NLP) technique that identifies latent semantic structures in large document collections \cite{abdelrazek2023topic}. The main idea is that, in patents, different semantic domains (identified through topics) correspond to different technological domains \cite{momeni2016identification}. While earlier studies often relied on Latent Dirichlet Allocation (LDA) for detecting topics in text, recent advances have demonstrated the superiority of transformer-based models \cite{najmani2023bertopic, gan2024experimental}. Specifically, we use BERTopic \cite{grootendorst2022bertopic}, a state-of-the-art approach that integrates contextual language models, statistical weighting, and clustering. BERTopic leverages BERT (Bidirectional Encoder Representations from Transformers) to generate dense sentence embeddings that capture the contextual meaning of each patent abstract \cite{devlin-etal-2019-bert}. These semantic representations are then grouped using clustering algorithms, which organize similar documents into coherent topical clusters. To interpret each cluster, BERTopic applies Term Frequency–Inverse Document Frequency (TF-IDF), a technique that highlights the most informative and distinctive words within each group. This combination allows for the extraction of coherent and interpretable topics, and has shown strong performance across scientific, industrial, and technological corpora \cite{guizzardi2023modeling, yun2024analysis, kim2024interdisciplinary, hou2025topic}. To date, no study has yet applied it to Green AI, as existing research relies on the more limiting standard Cooperative Patent Classification (CPC) system \cite{biggi2025green}. Based on the domains' identification, we assess the domains' impact by analyzing both their forward citations, a widely used proxy for the technological influence of patents on subsequent inventions \cite{trajtenberg1990penny, hall2000market}, and their market value, estimated through indicators of the stock market reaction to the patents' publication \cite{KoganPapanikolaouSeruStoffman2017}. We conclude by discussing the policy implications of our findings. 



\section*{Results}

First, we analyze the temporal development of Green AI patenting, examining trends in the number of filings over time. Second, we investigate the main assignees, focusing on their evolution, the concentration of patenting activity through top-10 shares and Gini index, and the geographic distribution of assignees at both the international level and within the United States. Third, we examine the technological content of the dataset by extracting CPC green subclasses and AI functional categories. Fourth, we apply BERTopic to identify sixteen distinct thematic clusters of technological domains. Fifth, by leveraging BERTopic's built-in UMAP dimensionality reduction, we explore the semantic similarity between these domains and identify four broader macro-domains. Finally, we analyze the temporal trajectories of all the domains we have identified and assess their impact in terms of forward citations and stock market value.

\subsection*{Temporal dynamics of Green AI patenting}

\begin{figure}
    \centering
    \includegraphics[width=0.75\linewidth]{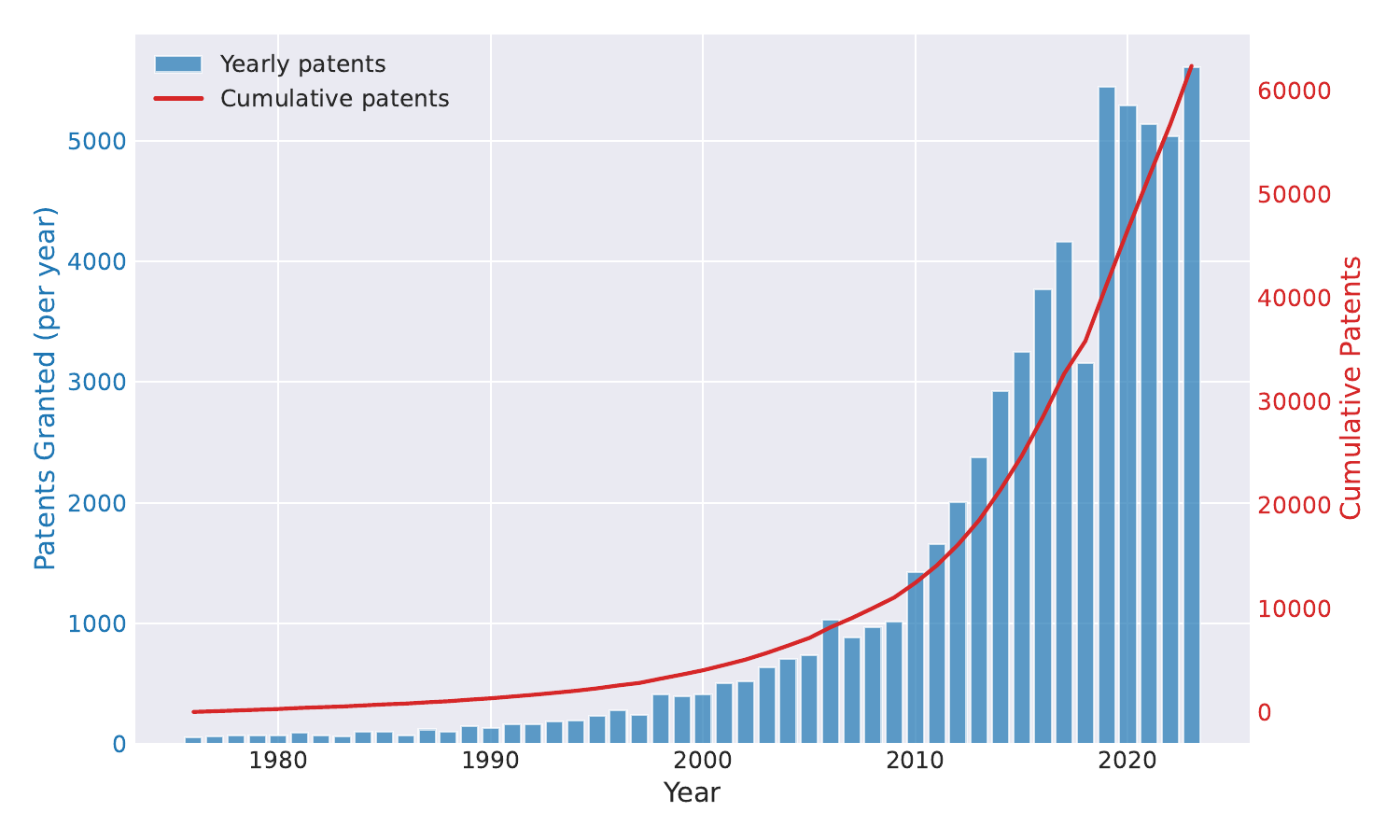}
    \caption{Development over time of the number of Green AI patents, both yearly and cumulatively, 1976-2023.}
    \label{fig:time-development}
\end{figure}

The trend of granted Green AI patents over time is depicted in Figure \ref{fig:time-development}. 
We see a modest yet relatively steady growth from the early 1970s, when only a few dozen patents were published each year, through the late 1990s, when annual publications surpassed one hundred. A significant increase occurred just before the year 2000, particularly in 1998, when the number of filings nearly doubled. This surge coincided with the Dot-Com Bubble, which led to a sharp rise in the stock market value of ICT-related companies in the late 1990s, driven by strong optimism around technological progress. This optimism, however, collapsed with the bursting of the bubble in 2000–2001 \cite{morris2012}.

Interestingly, the number of patents slightly declined in the year 2000— foreshadowing the impending crash. The following years, however, saw a strong rebound in patent activity, peaking in 2006 (with a brief exception in 2007). A marked recovery occurred again in 2010. Research confirms that the 2008 financial crisis heightened uncertainty, with heavy but short-term consequences on firms’ innovation and R\&D activity \cite{teplykh2017innovations}. This is perfectly consistent with studies highlighting the procyclical nature of patents, which tend to decline during economic downturns and rise in periods of recovery \cite{hingley2017business}.

Since 2010, the number of Green AI patents has grown continuously and significantly. Interestingly, this trend seems largely unaffected by the U.S. Supreme Court ruling Alice Corp. v. CLS Bank International (573 U.S. 208, 2014), which held that certain types of software-based inventions are not patentable. Although this decision introduced considerable uncertainty and led to the rejection of many AI-related applications, its effects appear to have been rapidly absorbed by the innovation system
\cite{whalen2022innovating}.

The growth trend, however, came to a halt in 2018, when the number of Green AI patents fell by nearly 20\%. While a decline is also observed in overall U.S. patenting activity, it was far less pronounced. We may hypothesize that this sector-specific downturn was influenced by trade tensions between the U.S. and China, which escalated during the first Trump administration. This decline, followed by a recovery beginning in 2019, aligns with empirical findings showing that trade disputes often suppress innovation in the short term but later stimulate increased R\&D investment, resulting in a U-shaped trajectory \cite{hu2025technological}.

Patent filings peaked in 2019 and then declined moderately in the following years, likely due to the COVID-19 pandemic and the associated uncertainty and strategic realignments. In 2023, the latest year for which data is available, Green AI patenting reached a new all-time high.

\subsection*{Patent Ownership and Concentration}

Analyzing assignee activity over the entire period highlights which organizations have made the largest cumulative investments in Green AI. However, aggregate rankings alone can obscure important shifts in leadership and evolving technological priorities. Table~\ref{tab:assignee_windows_full} divides the timeline into five-year windows, revealing distinct waves of innovation.
In the late 1980s, industrial control firms such as Westinghouse and Fanuc dominated early Green AI activity. During the 1990s, companies like Bosch and Intel emerged as key players, reflecting growing attention to energy-efficient electronics and semiconductor technologies. The 2000s marked a transition to electrification in the automotive sector and data-center optimization, with Honda, Ford, and Intel leading patenting efforts. In the most recent period (2015–2020), the focus has shifted decisively toward AI hardware and edge computing. Intel and IBM remained central actors, while Samsung and Qualcomm significantly expanded their presence. Over time, automotive, electronics, and industrial engineering firms have taken turns driving Green AI innovation, with a clear evolution from manufacturing control systems to computing architectures.

We evaluate market concentration with two distinct metrics: the share of the top ten assignees and the Gini coefficient. The Gini coefficient is a standard measure of inequality, originally developed to assess income distribution but widely applied to innovation studies to quantify concentration \cite{farris2010gini}. It ranges from 0 to 1, where 0 represents perfect equality (all assignees have the same number of patents) and 1 indicates maximum inequality (a single assignee holds all patents). The top-10 assignee share continuously decreased from 34.2\% (1985–1990) to 22.6\% (2015-2020). This suggests that in earlier periods, a small group of firms accounted for a larger fraction of total patents, and more recently this has developed into a relative broadening of participation at the top. In contrast, the Gini coefficient shows a steadily increasing trend (from 0.4761 in 1985–1990 to 0.7282 in 2015–2020), pointing to a growing disparity in patent ownership across the full distribution of assignees.

This apparent paradox can be explained by the increasing long-tail structure of the innovation ecosystem. While the top ten firms are now contributing a smaller share, the rest of the landscape is increasingly fragmented, with a large number of organizations holding only a few patents each. The Gini coefficient, being sensitive to the entire distribution (not just the top ranks), captures this widening gap between prolific assignees and the broader population of occasional contributors. In other words, Green AI patenting is becoming less top-heavy in terms of elite dominance but more uneven overall, as a few firms continue to scale up their portfolios disproportionately while many others remain peripheral players.

In terms of geography, the top assignees span key innovation-driven economies: Japan (e.g., Honda, Denso), the United States (IBM, Intel, Ford, Qualcomm), South Korea (Samsung, Hyundai), and Germany (Bosch, Siemens). Notably, Japan’s early dominance, driven by automotive and electronics firms, has gradually given way to greater U.S. leadership in AI computing hardware.

\begin{table}[ht]
\centering
\scriptsize
\resizebox{\linewidth}{!}{%
\begin{tabular}{@{} l HHHH lrrl lrrl lrrl lrrl @{}}
\toprule
& \multicolumn{4}{H}{\textbf{Overall}} 
& \multicolumn{4}{c}{\textbf{1985--1990}} 
& \multicolumn{4}{c}{\textbf{1995--2000}} 
& \multicolumn{4}{c}{\textbf{2005--2010}} 
& \multicolumn{4}{c}{\textbf{2015--2020}} \\
\cmidrule(r){6-9} \cmidrule(l){10-13} \cmidrule(l){14-17} \cmidrule(l){18-21}
\textbf{Metric} 
& \multicolumn{4}{H}{Gini: 0.8 Top-10: 20.9\%}
& \multicolumn{4}{c}{Gini: 0.476 Top-10: 34.2\%}
& \multicolumn{4}{c}{Gini: 0.543 Top-10: 29.1\%}
& \multicolumn{4}{c}{Gini: 0.657 Top-10: 26.7\%}
& \multicolumn{4}{c}{Gini: 0.728 Top-10: 22.6\%} \\
\addlinespace
\textbf{Rank} 
& \textbf{Assignee} & \textbf{Cnt} & \textbf{\%} & \textbf{Ctry}
& \textbf{Assignee} & \textbf{Cnt} & \textbf{\%} & \textbf{Ctry}
& \textbf{Assignee} & \textbf{Cnt} & \textbf{\%} & \textbf{Ctry}
& \textbf{Assignee} & \textbf{Cnt} & \textbf{\%} & \textbf{Ctry}
& \textbf{Assignee} & \textbf{Cnt} & \textbf{\%} & \textbf{Ctry} \\
\cmidrule(r){6-9} \cmidrule(l){10-13} \cmidrule(l){14-17} \cmidrule(l){18-21}
1  & Honda     & 2677 & 4.3\% & JP & Honda         & 90  & 14.6\% & JP & Honda     & 198 & 10.9\% & JP & Honda     & 398 & 6.8\% & JP & Intel     & 983 & 3.9\% & US \\
2  & IBM       & 2101 & 3.4\% & US & Westinghouse  & 29  & 4.7\%  & US & IBM       & 69  & 3.8\%  & US & IBM       & 239 & 4.1\% & US & IBM       & 939 & 3.8\% & US \\
3  & Intel     & 1985 & 3.2\% & US & Fanuc         & 19  & 3.1\%  & JP & Bosch     & 38  & 2.1\%  & DE & Intel     & 209 & 3.6\% & US & Honda     & 738 & 3.0\% & JP \\
4  & GE        & 1212 & 2.0\% & US & GE            & 13  & 2.1\%  & US & Intel     & 36  & 2.0\%  & US & Ford      & 132 & 2.2\% & US & Ford      & 574 & 2.3\% & US \\
5  & Ford      & 1166 & 1.9\% & US & Denso         & 13  & 2.1\%  & JP & GE        & 36  & 2.0\%  & US & AMD       & 115 & 2.0\% & US & Samsung   & 532 & 2.1\% & KR \\
6  & Samsung   & 1056 & 1.7\% & KR & Bosch         & 12  & 2.0\%  & DE & Fuji Elec & 35  & 1.9\%  & JP & GE        & 112 & 1.9\% & US & GE        & 531 & 2.1\% & US \\
7  & GM        & 834  & 1.3\% & US & Fuji Elec     & 9   & 1.5\%  & JP & Siemens   & 34  & 1.9\%  & DE & Siemens   & 102 & 1.7\% & DE & Qualcomm  & 347 & 1.4\% & US \\
8  & Apple     & 630  & 1.0\% & US & Boeing        & 9   & 1.5\%  & US & Motorola  & 31  & 1.7\%  & US & GM        & 89  & 1.5\% & US & GM        & 345 & 1.4\% & US \\
9  & Hyundai   & 623  & 1.0\% & KR & IBM           & 8   & 1.3\%  & US & AMD       & 30  & 1.6\%  & US & Rockwell  & 87  & 1.5\% & US & Apple     & 339 & 1.4\% & US \\
10 & Siemens   & 614  & 1.0\% & DE & GM            & 8   & 1.3\%  & US & TI        & 24  & 1.3\%  & US & Bosch     & 83  & 1.4\% & DE & Hyundai   & 322 & 1.3\% & KR \\
\bottomrule
\end{tabular}
}
\caption{Top 10 patent assignees and concentration metrics across time windows (1985--1990, 1995--2000, 2005--2010, 2015--2020). Gini coefficients and top-10 shares measure concentration over time.}
\label{tab:assignee_windows_full}
\end{table}

\subsection*{Geographical Distribution}

\begin{figure}[H]
    \centering
    \includegraphics[width=0.9\linewidth]{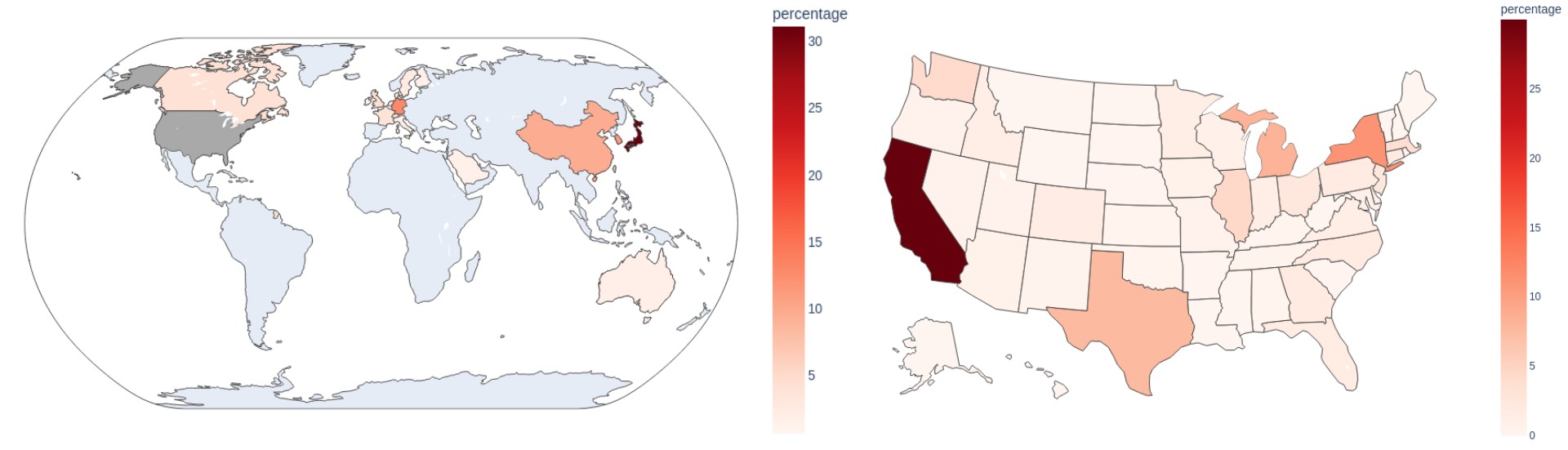}
    \caption{(left) Country-level distribution of Green AI patents, using assignee location as a proxy. The United States is obscured in grey to better visualize activity in other countries. The heatmap reflects the relative share of non-US patents across countries. (right) US-level distribution of Green AI patents, using assignee location as a proxy.
}
    \label{fig:assignee_maps}
\end{figure}

Considering the assignee location as a proxy of the geographical distribution, patents exhibit a strong geographical concentration of activity. This is shown in Figure~\ref{fig:assignee_maps} (left). Data on the United States are omitted from Figure~\ref{fig:assignee_maps} (left), and shown at a finer level in Figure~\ref{fig:assignee_maps} (right). This is because the United States are overwhelmingly dominant, accounting for 59.3\textbf{\%} of all entries. However, this is not surprising, given that we are looking at patents granted by the United States Patent and Trademark Office. Other countries with notable contributions include Japan (12.7\textbf{\%}), Germany (5.1\textbf{\%}), South Korea (4.0\%), and China (3.7\%). European countries such as the United Kingdom (1.5\%), France (1.3\%), and Switzerland (1.1\%) also appear, albeit with smaller shares. 

At the subnational level within the United States, Figure~\ref{fig:assignee_maps} (right) shows that the distribution is  skewed as well. California dominates with nearly 30\% of US-based entries, followed by New York (11.1\%) and Michigan (8.2\%). This reflects the strength of California’s high-tech clusters \cite{Klepper2010} and Detroit’s automotive industry \cite{Mattoon2014}. Other key contributors include Texas (7.6\%), Illinois (4.4\%), Washington (4.2\%), and Massachusetts (3.8\%), all hosting major industrial or tech hubs.

In conclusion, the country-level map highlights the strong presence in East Asia and Europe alongside the US, while the US state-level heatmap reveals a clear clustering along the coasts and in key industrial corridors. To deepen the analysis of international leadership in Green AI, we build country-level collaboration networks, both at the organizational level (assignees) and the individual level (inventors), which are provided in the Supplementary Material (Table ST1 and Figure SF1).

\subsection*{CPC Green Subclasses and AI Functional Categories}

We investigate the technological content of our patent dataset on both green and AI fronts (Table~\ref{tab:combined_cpc_ai}). For green technologies, we rely on the Cooperative Patent Classification (CPC) subclasses under the \texttt{Y02} and \texttt{Y04S} schemes, which refine the broader CPC category of ‘green’ patents and highlight specific technological domains \cite{Angelucci2018}.

The most common subclass is \texttt{Y02T} (Transportation mitigation, 23.5\%), underscoring the central role of sustainable mobility. This is followed by \texttt{Y02D} (ICT energy reduction, 18.5\%) and \texttt{Y02E} (Energy generation and distribution, 14.7\%), indicating a focus on energy optimization. Other key categories include \texttt{Y02P} (Manufacturing mitigation) and \texttt{Y04S} (Smart grids), both over 12\%.
Less frequent categories such as \texttt{Y02B} (Buildings and appliances), \texttt{Y02A} (Climate adaptation), \texttt{Y02W} (Waste/wastewater), and \texttt{Y02C} (GHG capture) suggest more specialized or emerging uses.

On the AI side, predictive labels from the Artificial Intelligence Patent Dataset (AIPD)\cite{pairolero2024aipd} show that AI-Assisted Planning and Control dominates (57.7\%), pointing to strong industrial applications. AI Hardware follows (38.6\%), indicating important patenting activity in hardware-level innovation, likely aimed at improving the energy efficiency or capabilities of devices. Knowledge Representation and Reasoning (18.4\%) and Computer Vision (10.2\%) are also relevant, especially for automation. Traditional Machine Learning (9.3\%) and Evolutionary Computation (8.3\%) contribute meaningfully, while NLP and Speech Processing remain niche.

\begin{table}[ht]
\centering
\resizebox{0.9\textwidth}{!}{%
\begin{tabular}{lrr  lrr}
\toprule
\multicolumn{3}{c}{\textbf{CPC Subclass (Y02–Y04S)}} & \multicolumn{3}{c}{\textbf{AI Functional Category (AIPD)}} \\
\cmidrule(l){1-3} \cmidrule(l){4-6}
\multicolumn{1}{c}{\textbf{Subclass}} & \textbf{Count} & \textbf{Pct.\ (\%)} 
  & \multicolumn{1}{c}{\textbf{Category}} & \textbf{Count} & \textbf{Pct.\ (\%)} \\
\cmidrule(l){1-3} \cmidrule(l){4-6}
Y02T: Transportation mitigation          & \np{23614} & 23.5 
  & AI-Assisted Planning \& Control       & \np{36556} & 57.7 \\
Y02D: ICT energy reduction               & \np{18637} & 18.5 
  & AI Hardware                           & \np{24432} & 38.6 \\
Y02E: Energy generation \& distribution  & \np{14749} & 14.7 
  & Knowledge Representation \& Reasoning & 11\,640    & 18.4 \\
Y02P: Manufacturing mitigation           & 14\,563    & 14.5 
  & Computer Vision                       &  6\,451    & 10.2 \\
Y04S: Smart grids                        & 12\,820    & 12.8 
  & Machine Learning                      &  5\,893    &  9.3 \\
Y02B: Buildings \& appliances mitigation &  8\,521    &  8.5 
  & Evolutionary Computation              &  5\,275    &  8.3 \\
Y02A: Climate change adaptation          &  6\,019    &  6.0 
  & Natural Language Processing           &  3\,143    &  5.0 \\
Y02W: Waste \& wastewater management     &  1\,377    &  1.4 
  & Speech Processing                     &  1\,100    &  1.7 \\
Y02C: GHG capture \& storage             &    202     &  0.2 
  &                                        &           &      \\
\bottomrule
\end{tabular}%
}
\caption{Green CPC Subclasses and AI Functional Categories (AIPD)}
\label{tab:combined_cpc_ai}
\end{table}

\subsection*{Discovery of technological domains with Topic Modeling}
The \texttt{BERTopic} model identified sixteen distinct topics within the Green AI patent corpus. In our framework, these topics are interpreted as \textbf{technological domains}, which means coherent areas of innovation characterized by shared semantic and functional features. More details on the theoretical foundations of the model and the rigorous selection of its hyperparameters are provided in the Methods section. To ensure interpretability, we manually assigned a descriptive label to each topic, reflecting its underlying technological domain. These labels were derived through qualitative inspection of the top-ranked keywords generated by the model, their class-based probabilities, and close reading of representative patent abstracts within each domain. 
Table \ref{tab:topics} provides an overview of all sixteen domains, including (i) the assigned thematic label, (ii) the most representative keywords and their relative importances, and (iii) the number of patents associated with each domain. Several key patterns emerge from this mapping. 
{Domain~0}, labeled \textit{Data Processing \& Memory Management}, dominates the corpus with over \np{27000} documents. High-probability keywords such as \textit{processing} (0.516), \textit{computing} (0.461), and \textit{memory} (0.421) suggest a strong emphasis on computing systems research, encompassing areas like algorithmic optimization, memory hierarchies, operating system architecture, and database design.

A significant cluster of energy-related research is captured by {Domains~1} (\textit{Microgrid \& Distributed Energy Systems}), {2} (\textit{Vehicle Control \& Autonomous Powertrains}), {4} (\textit{Photovoltaic \& Electrochemical Devices}), and {7} (\textit{Battery Charging \& Management}), collectively accounting for approximately \np{14000} documents. These domains show a strong orientation toward sustainable energy technologies, including smart grid integration, electric vehicles, power electronics, and battery systems. 

Environmental monitoring and resource management emerge in {Domains~3} (\textit{Irrigation \& Agricultural Water Management}, \np{2754} docs) and {13} (\textit{Meteorological Radar \& Weather Forecasting}, \np{542}), with an applied focus in agricultural engineering and climate instrumentation. 

\begin{table}[H]
\centering
\small
\resizebox{\linewidth}{!}{%
\begin{tabular}{c l p{11.3cm} r H}
\toprule
\textbf{Id} & \multicolumn{1}{c}{\textbf{Domain (labeled)}} & \multicolumn{1}{c}{\textbf{Top Keywords (prob)}} & \textbf{N. Patents} & \textbf{Macro-Domain} \\
\midrule
0 & Data Processing \& Memory Management            & processing (.52), computing (.46), process (.45), systems (.44), memory (.42)       & \np{27435} & 4 \\
1 & Microgrid \& Distributed Energy Systems         & microgrid (.49), electricity (.42), utility (.40), power (.38), energy (.37)         &  \np{5378} & 3 \\
2 & Vehicle Control \& Autonomous Powertrains       & vehicle (.48), vehicles (.47), control (.42), driving (.39), engine (.39)            &  \np{3747} & 1 \\
3 & Irrigation \& Agricultural Water Management     & irrigation (.51), systems (.43), flow (.35), process (.35), water (.33)             &  \np{2754} & 4 \\
4 & Photovoltaic \& Electrochemical Devices         & semiconductor (.52), photoelectric (.51), electrodes (.49), electrode (.47), photovoltaic (.47) &  \np{2599} & 4 \\
5 & Clinical Microbiome \& Therapeutics             & microbiome (.48), clinical (.37), physiological (.32), therapeutic (.32), disease (.31)        &  \np{2286} & 4 \\
6 & Combustion Engine Control                       & combustion (.42), engine (.37), control (.34), fuel (.34), ignition (.32)            &  \np{2179} & 1 \\
7 & Battery Charging \& Management                  & charging (.49), charger (.45), charge (.42), battery (.39), batteries (.38)          &  \np{1541} & 3 \\
8 & HVAC \& Thermal Regulation                      & hvac (.52), heater (.47), cooling (.47), heating (.46), evaporator (.45)             &  \np{1523} & 3 \\
9 & Lighting \& Illumination Systems                & lighting (.62), illumination (.60), lights (.54), brightness (.53), light (.49)      &  \np{1219} & 4 \\
10 & Exhaust \& Emission Treatment                  & exhaust (.46), catalytic (.45), purification (.44), catalyst (.37), emissions (.36)  &  \np{1064} & 1 \\
11 & Wind Turbine \& Rotor Control                  & turbines (.50), turbine (.49), windmill (.46), wind (.42), rotor (.30)              &   988 & 3 \\
12 & Aircraft Wing Aerodynamics \& Control          & wing (.45), aircraft (.45), wingtip (.42), apparatus (.42), aerodynamic (.42)       &   697 & 2 \\
13 & Meteorological Radar \& Weather Forecasting    & radar (.54), meteorological (.51), weather (.41), precipitation (.39), systems (.37)&   542 & 3 \\
14 & Fuel Cell Systems \& Electrodes                & fuel (.37), cell (.31), systems (.29), cells (.29), controls (.26)                  &   377 & 1 \\
15 & Turbine Airfoils \& Cooling                    & airfoils (.58), airfoil (.57), turbine (.43), engine (.33), axial (.32)              &   352 & 2 \\
\bottomrule
\end{tabular}
}
\caption{Identified technological domains with their most common words (with probability of appearing in that domain)  and their number of patents.}
\label{tab:topics}
\end{table}

Conventional propulsion technologies remain well-represented through {Domain~6} (\textit{Combustion Engine Control}) and {Domain~10} (\textit{Exhaust \& Emission Treatment}), with nearly \np{3300} documents covering internal combustion optimization and emission reduction. {Domains~11} (\textit{Wind Turbine \& Rotor Control}) and {15} (\textit{Turbine Airfoils \& Cooling}) mirror the importance of aerodynamics and power generation. 

Energy-efficient building technologies are also important, with {Domains~8} (\textit{HVAC \& Thermal Regulation}) and {9} (\textit{Lighting \& Illumination Systems}). {Domain 5} (\textit{Clinical Microbiome \& Therapeutics}) reflects innovations at the intersection of biomedical science and AI, that are increasingly relevant in the context of climate-related research. In fact, recent studies highlight the environmental footprint of healthcare \cite{lenzen2020environmental} and pharmaceutical pollution \cite{aus2016pharmaceuticals}, that can be addressed by AI-driven advances in precision medicine and microbiome research. Differently, {Domain~14} (\textit{Fuel Cell Systems \& Electrodes}) is rooted in material science and energy technologies, focusing on electrochemical systems such as hydrogen fuel cells. 

Lastly, aerodynamics is further addressed by {Domain~12} (\textit{Aircraft Wing Aerodynamics \& Control}) and again by {Domain~15}, indicating a split across applications in aviation and turbomachinery.

Overall, this taxonomy reveals significant thematic cores across computing, energy, environment, transport, and materials science.


\subsection*{Technological (Macro-)Domains and Their Temporal Dynamics}
\begin{figure}[h]
    \centering
    \includegraphics[width=1\linewidth]{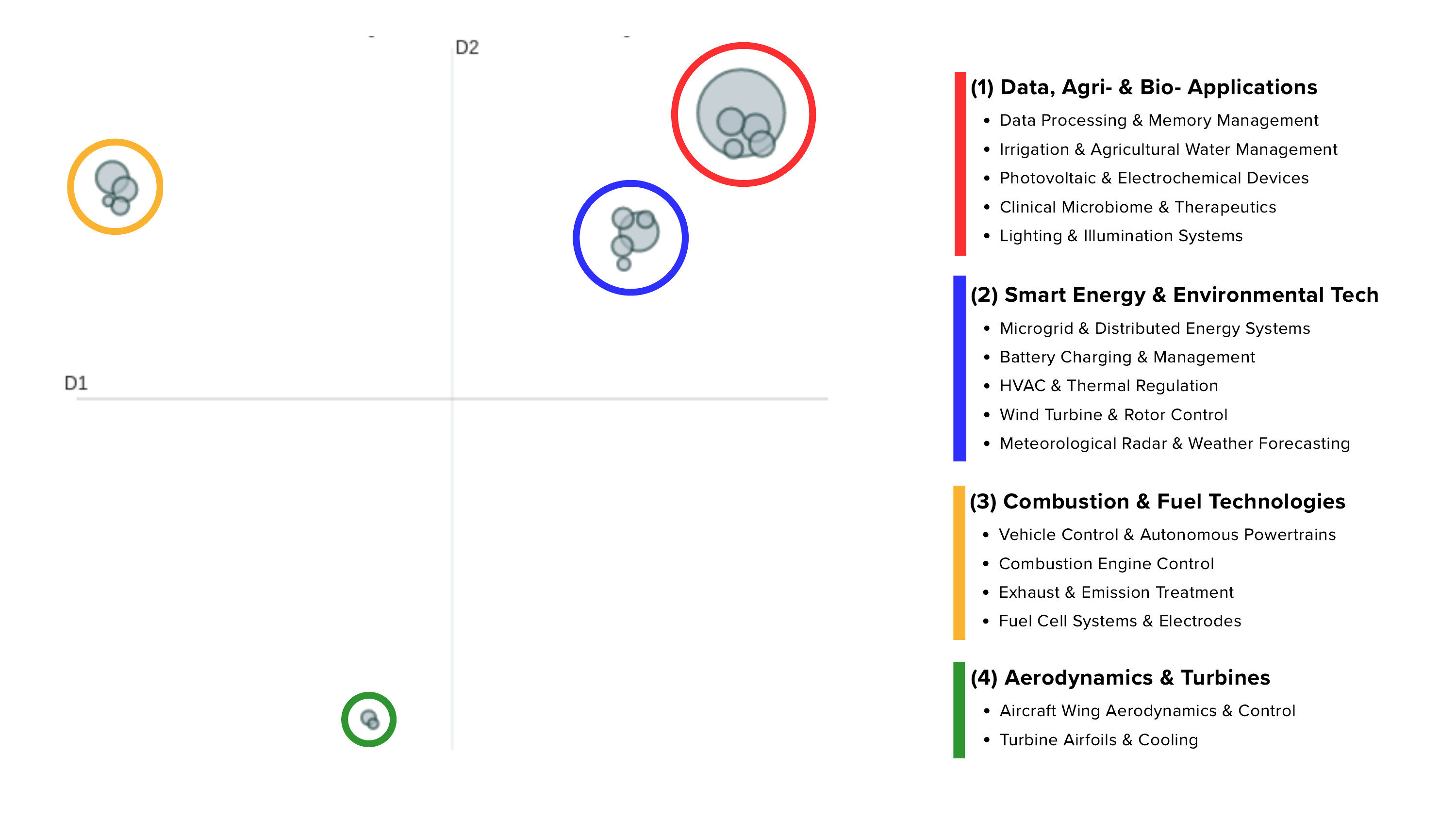}
    \caption{Projection in 2D-space of technological domains derived from BERTopic through dimensionality reduction (UMAP). Each grey bubble represents a single technological domain. The size of the bubbles reflects the domain importance in terms of number of patents. Each colored circle constitutes a manually labeled macro-domain, aggregating domains characterized by proximity in the Euclidean space. We display the four macro-domains, along with their associated technological domains, to illustrate the thematic structure and internal composition of Green AI innovation.}
    \label{fig:umap}
\end{figure}

To further explore the semantic relationships between domains, we leverage \texttt{UMAP} (Uniform Manifold Approximation and Projection), a dimensionality reduction technique built into the BERTopic framework. UMAP projects the high-dimensional embeddings  
derived from the BERT-based representations, into a two-dimensional space, allowing for intuitive visualization \cite{grootendorst2022bertopic}. In particular, one typically focuses on the  centroids of the topics (technological domains in this paper). 
The resulting plot (see Figure~\ref{fig:umap}) displays each domain as a grey bubble centered in the centroid of the domain and with a radius proportional to the number of patents in the domain. In such representation, spatial proximity reflects semantic similarity: domains that appear closer together tend to share more lexical and conceptual features  \cite{grootendorst2022bertopic}. We note that Figure~\ref{fig:umap} reveals the emergence of four well-defined groups of related domains, each corresponding to a coherent thematic area within the corpus ({macro-domain}).

Macro-Domain 1 - \textbf{\textit{Data, Agri- \& Bio-Applications}} (Red, upper-right quadrant), is the most expansive both in document count  and semantic variance. 
The co-location of terms such as \textit{processing}, \textit{irrigation}, \textit{photovoltaic}, \textit{microbiome}, and \textit{lighting} reveals a wide-reaching intersection of data-intensive methods applied across diverse sectors: agritech, biotech, smart materials, lighting systems, and more.

Macro-Domain 2 – \textbf{\textit{Smart Energy \& Environmental Tech}} (Blue, upper-right quadrant) brings together areas focused on renewable energy infrastructure, building automation, and climate monitoring. The convergence of keywords like \textit{grid}, \textit{battery}, \textit{heater}, \textit{ventilation}, and \textit{radar} highlights the integration of distributed energy systems, storage solutions, environmental sensing, and smart control technologies.

Macro-Domain 3 - \textit{\textbf{Combustion \& Fuel Technologies}} (Orange, upper-left quadrant), is composed of domains focused on onboard energy conversion and emission systems. 
Their spatial proximity indicates shared terminology such as \textit{engine}, \textit{combustion}, \textit{fuel}, and \textit{catalytic}.

Macro-Domain 4 – \textbf{\textit{Aerodynamics \& Turbines}} (Green, lower-left quadrant) is characterized by concepts related to high-speed fluid dynamics, such as \textit{airfoil}, \textit{blade}, and \textit{cooling}. It reflects a focus on aerodynamic performance and thermal regulation in turbine and aerospace applications, and is spatially distinct from combustion-related research.

\begin{figure}[h]
    \centering
    \vspace{-0.3cm}
    \includegraphics[width=1.0\linewidth]{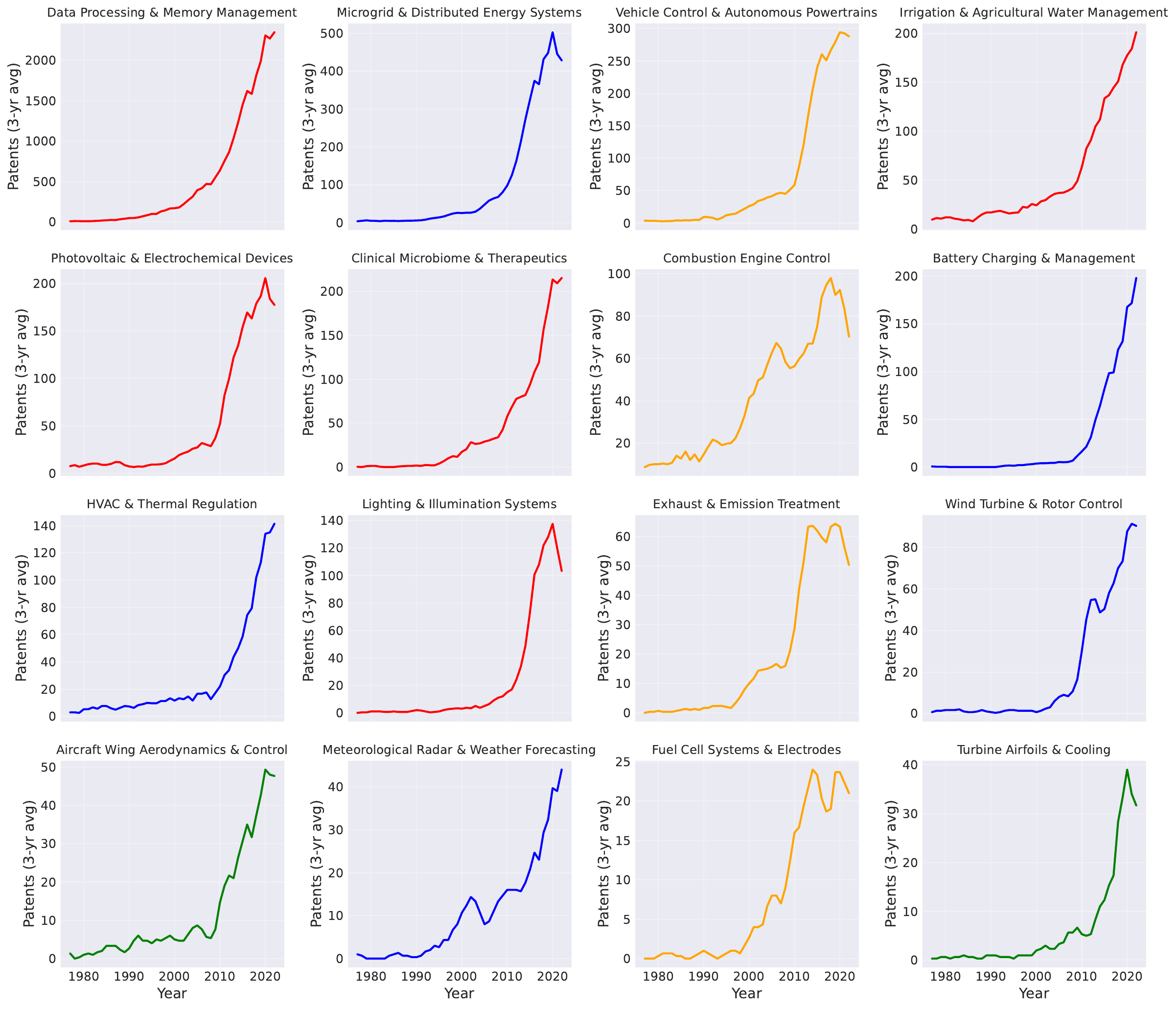}
    \vspace{-0.3cm}
    \caption{Temporal evolution of patenting activity for each technological domain, displayed as individual line charts with a 3-year rolling average. Each panel illustrates the number of granted patents per year for a specific domain, capturing trends in innovation intensity from 1976 to 2023. The color of each line corresponds to the macro-domain to which the technological domain belongs as per Figure~\ref{fig:umap}. 
    }
    \label{fig:individual-evolution}
\end{figure}

We now examine the temporal evolution of the identified domains and macro-domains.
In Figure~\ref{fig:individual-evolution} we display the temporal evolution of each domain individually, using a centered 3-year rolling average of granted patents from 1976 to 2023. To maintain continuity with the discovered macro-domains, each domain's trend is color-coded according to its associated macro-domain.


Several distinct
patterns emerge. Sustained growth is evident in domains such as 
\textit{Data Processing \& Memory Management}, \textit{Microgrid \& Distributed Energy Systems}, \textit{Battery Charging \& Management}, and \textit{Photovoltaic \& Electrochemical Devices}, all of which 
show continuous upward trajectories into the 2020s. Other domains, such as \textit{Clinical Microbiome \& Therapeutics}, \textit{Irrigation \& Agricultural Water Management}, and \textit{Wind Turbine \& Rotor Control}, have
emerged more recently but exhibit similar strong momentum.
In contrast, plateauing or declining trends are observed in legacy domains. \textit{Combustion Engine Control} and \textit{Exhaust \& Emission Treatment} both peaked around the mid-to-late 2010s and have since leveled off or declined. \textit{Lighting \& Illumination Systems}, \textit{Fuel Cell Systems \& Electrodes}, and \textit{Turbine Airfoils \& Cooling} show comparable downturns, likely indicating technological maturity or shifting innovation priorities.

This decomposition highlights the rise of data-intensive and clean energy domains while marking a relative decline in traditional combustion and emission-related technologies.

\subsection*{Impact of Green AI Technological Domains on Knowledge Flows and Market Value}

We combine two complementary measures of technological importance per domain: 
(i) the average 
forward citations of patents, 
a proxy for the extent to which subsequent inventions build upon a given domain’s knowledge base, and (ii) the mean market value of innovation, measured in 1982-equivalent million USD, which captures the average private economic value of patents as inferred from the stock market reaction to patent grant announcements. These data are integrated from the study of Kogan et al. \cite{KoganPapanikolaouSeruStoffman2017}, and in particular from their most recent (2023) data release.
Forward citations are widely used in the innovation literature as an objective indicator of “knowledge flow” and technological influence across later patents \cite{AristodemouTietze2018}, while market value is related to the private economic returns on R\&D investments by publicly listed entities. Since not all patents in our dataset are assigned to publicly listed firms, we have market value data for approximately 42.2\% of the patents, but there is no reason to believe that the value of patents held by unlisted firms should behave any differently at the micro- and macro-domains level.

\begin{figure}[h]
    \centering
    \includegraphics[width=1\linewidth]{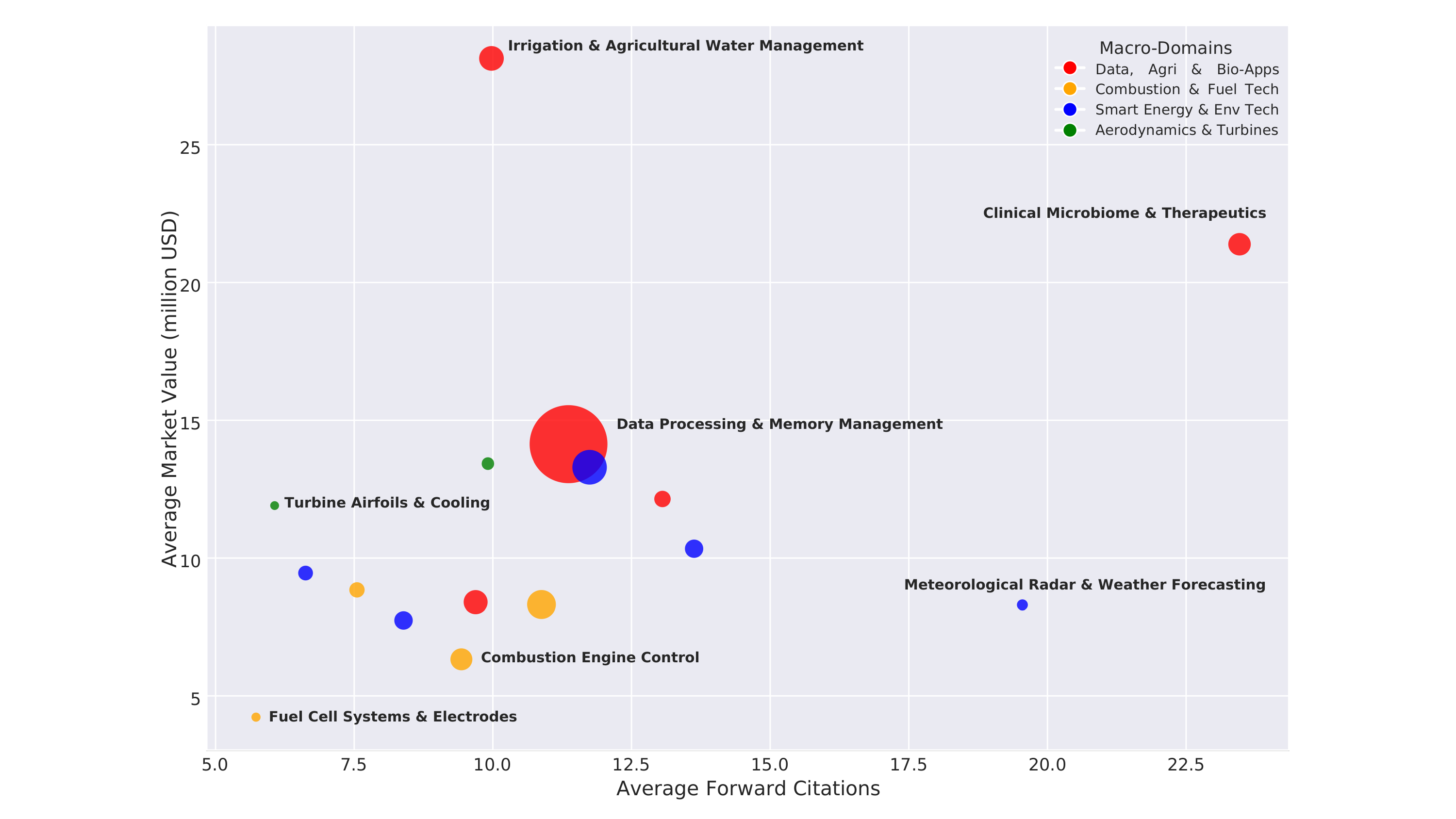}
\vspace{-0.4cm}
    \caption{Scatter plot of average forward citations versus average market‐value (deflated to 1982 million USD) per patent for sixteen Green AI technological domains, sized by the number of patents in each domain and colored by macro‐domain. Marker size scales with topic patent volume (largest for Data Processing \& Memory Management), while position reflects technological influence (horizontal) and private economic returns (vertical). We highlight some highest- and least‐impact topics to illustrate the upper and lower end of both citation and market‐value distributions.
}
    \label{fig:impact}
\end{figure}

The scatterplot shown in Figure~\ref{fig:impact} presents the relationship between the average number of forward citations and the average market value across all identified technological domains. Each bubble represents a domain, with its size proportional to the number of patents and its color indicating its macro-domain.

The domain \textit{Clinical Microbiome \& Therapeutics} emerges as a clear outlier, exhibiting both the highest citation intensity and the highest market value. A similarly notable result is observed for\textit{ Irrigation \& Agricultural Water Management}, which ranks among the highest in terms of market value despite only moderate citation levels, highlighting the strong commercial potential of agricultural and environmental technologies.

The domain \textit{Data Processing \& Memory Management} stands out for its large patent volume, moderate citation rate, and relatively high market value. This reflects the central role of efficient computing in energy-conscious innovation, and in general broad relevance and scalability of computing technologies across industries.

Interestingly, the figure also reveals domains where scientific impact and market value diverge. For instance, \textit{Meteorological Radar \& Weather Forecasting} is highly cited but has relatively modest market value, suggesting that while this domain generates considerable academic or technological attention, it may yield less direct economic benefit to firms. 
Conversely, \textit{Turbine Airfoils \& Cooling} achieves above-average market value despite a relatively low citation count.

In contrast, patents related to the macro-domain of\textit{ Combustion \& Fuel Technologies} tend to show modest citation and market value levels. These areas likely reflect more mature or legacy technologies, where innovation continues but without generating high-impact breakthroughs or strong commercial returns.

Overall, forward citations and market value offer complementary perspectives. Domains at the intersection of data, agriculture, and energy (particularly those in the red and, and in part blue macro-domains) tend to perform well across both dimensions, whereas traditional propulsion technologies show signs of stagnation in both influence and value.




 

\section*{Discussion}


This study investigates the intersection of artificial intelligence and climate technologies by analyzing a newly constructed dataset of approximately \num{63000} Green AI patents. These are patents that are both classified under climate-related CPC categories (Y02 and Y04S) and identified as AI-related using the Artificial Intelligence Patent Dataset (AIPD)\cite{pairolero2024aipd}. A key contribution of this research is the release of this Green AI patent dataset, which offers a valuable resource for future studies on sustainable innovation and AI policy.

The temporal analysis of Green AI patenting reveals a trajectory shaped by both macroeconomic and technological cycles, with notable increases during the ICT boom and consistent growth after 2010 despite major shocks.

Instead, the evolution of Green AI patent assignees over different time windows reveals a clear shift from traditional industrial control and combustion technologies toward data-centric and energy-efficient domains. This shift was accompanied by the growth of a larger innovation ecosystem, with a few players holding increasingly large patent portfolios. The geographical distribution of patents is highly uneven, with well-identified innovation hotsposts at the global level and across US States.  


We applied advanced NLP-techniques on patent abstracts, and in particular BERTopic, in order to discover the presence of underlying technological domains based on semantic patterns. Its results yielded sixteen domains, grouped in four macro-domains: \textit{Combustion \& Fuel Technologies, Aerodynamics \& Turbines, Smart Energy \& Environmental Tech}, and \textit{Data, Agri- \& Bio-Applications}. The temporal evolution of patenting activity for each topic reveals some important dynamics in Green AI innovation. First, significant growth post-2010 in topics like \textit{Data Processing \& Memory Management} and \textit{Microgrid \& Distributed Energy Systems}, which reflects the digitization of energy and climate systems. Overall, data management, energy distribution, and vehicle control are among the most patent-intensive domains, while agriculture, batteries, and photovoltaic systems are emerging with momentum. Legacy areas such as combustion engine control are stagnating, as a consequence of the transition away from fossil fuel-based innovation.

We also analyzed both the technological impact (in terms of forward citations) and the economic value (in terms of stock market reaction to the patent) across technological domains. Some of them, such as \textit{Clinical Microbiome \& Therapeutics} and \textit{Irrigation \& Agricultural Water Management}, demonstrate high market value, even with moderate citation intensity, indicating strong commercial applicability. Conversely, domains like \textit{Meteorological Radar} have high citations but low economic return. Domains related to the macro-domain of \textit{Combustion \& Fuel Technologies} are characterized by both low economic value and modest technological impact. 

These findings have policy relevance. First, it is important to monitor the rising industrial concentration levels of inventive activities, to prevent monopolistic behaviors in the exploitation of Green AI patenting. 
 Second, according to the impact analysis, some domains such as agriculture, data processing, and clean energy systems showed high commercial and technological momentum, and therefore are likely to attract continued or increased R\&D investment. Naturally, policies may prioritize, and therefore incentivize through government support, some domains and not others, and changes in the policy cycle can have dramatic effects on both innovation and the diffusion of technologies. Potentially vulnerable domains include those with lower market value but high social or environmental importance, such as meteorological monitoring and emission treatment. These areas may lack strong private incentives, but are essential for climate adaptation.

While this study provides valuable insights into the intersection of artificial intelligence and climate technologies through patent data, several limitations should be acknowledged.
First, the analysis is inherently limited by its reliance on patent records as a proxy for innovation. Not all technological advancements are patented: some remain as trade secrets, others are disclosed through publications, and some innovations, especially in software or data-centric fields, may not be patented at all due to strategic or legal considerations \cite{graham2003intellectual}. 

Second, this study focuses on patents granted by the United States Patent and Trademark Office (USPTO). While the U.S. is a major hub for technological innovation and a key continental market for the world economy, some patenting behaviors (including patentability conditions) might not be fully generalizable to other contexts, and complementary analyses could be conducted in further research on other patents systems (e.g. European Patent Office - EPO).


\section*{Conclusions}
Climate change is a major global challenge, and artificial intelligence (AI) offers promising opportunities to accelerate climate technologies. This study analyzes the contribution of AI to climate mitigation and adaptation by identifying key technological domains and trends in “Green AI” innovation using U.S. patent data.

The analysis maps historical dynamics, leading corporate actors (notably in automotive and electronics), and the geographical concentration of innovation (primarily in the U.S., especially California, with Japan, South Korea, and Germany as further important hubs). 

Using BERTopic, we identified sixteen technological domains, grouped in four macro-domains. Their temporal trajectories show a clear shift from legacy areas like combustion engines toward fast-growing fields such as data processing, microgrids, and agricultural water management.

In terms of impact, Green AI domains such as \textit{Clinical Microbiome \& Therapeutics} and \textit{Agricultural Water Management} exhibit high market value, while others like Data Processing \& Memory Management combine strong citation intensity with broad commercial potential. Others, such as \textit{Meteorological Radar \&  Weather Management} have high technological impact but modest commercial incentive, and could therefore be sustained by tailored policies.

By analyzing the structure, dynamics, and direction of AI-driven climate innovation, and by creating and releasing to the research community the Green AI patent dataset, this work contributes to a growing understanding a key area of technological progress, and helps build the foundations for novel strategic insights for researchers, firms, and policymakers interested in data-driven innovation policy.

\section*{Methods}


This study employs a five‐stage methodology to analyze the thematic composition of Green AI patents:
\begin{enumerate}
  \item Dataset construction  
  \item Descriptive analysis  
  \item Topic modeling with BERTopic, including hyperparameter tuning
  \item UMAP-based 2D representation, and analysis of the temporal evolution of technological domains
    \item Impact analysis with forward citations and stock market value
\end{enumerate}

\subsection*{Dataset Construction}

Our analysis is based on a patent dataset built by intersecting two main sources:

\begin{itemize}
    \item \textbf{PatentsView (USPTO):} This is the official repository of U.S. patents, containing data from 1976 onwards \cite{toole2021patentsview}. U.S. patents are widely regarded as a strong indicator of technological developments at the global level, particularly in high-tech and ICT sectors \cite{NBERw24793}. For our study, we rely on the version of PatentsView available as of January 2025. We extract all patents related to environmental technologies by adopting a widely accepted taxonomy from the literature \cite{Angelucci2018}, specifically the Cooperative Patent Classification (CPC) classes \texttt{Y02} (climate change mitigation and adaptation technologies) and \texttt{Y04S} (smart grids). These classes define the ``green'' component of our dataset.

    \item \textbf{Artificial Intelligence Patent Dataset (AIPD) \cite{pairolero2024aipd}, 2024:} This is the second edition of the AIPD, updated through 2023 and developed by the USPTO. The authors started from a dataset of U.S. patents and publications and applied a machine learning model to identify those containing AI-related components. The model uses a binary classification approach, trained on positive and negative examples based on text analysis of patent abstracts and claims, as well as citation information. The classification was manually validated by USPTO experts. 
    The resulting dataset includes, among other variables, an AI prediction score for each of eight AI technology categories: machine learning, evolutionary computation, natural language processing, speech, vision, knowledge processing, planning and control, and AI hardware. A patent is classified as AI-related if it receives a prediction score of at least 0.5 in at least one of these categories; in such cases, the variable \texttt{predict50\_any\_ai} is set to 1, otherwise to 0. The selected AI-related patents cover multiple application areas, among which climate-related technology, health, industrial applications, etc.
\end{itemize}

By intersecting the set of ``green'' patents with those identified by the AIPD as AI-related (i.e., with those having \texttt{predict50\_any\_ai = 1}), we obtain a final set of {\np{63326} ``Green AI'' patents}.



\subsection*{Descriptive Analysis}

To explore the characteristics of our Green AI dataset, we extracted key metadata including filing dates, assignee names and locations, abstracts, and data on the CPC subclasses.  These variables allow us to analyze the temporal and geographical distribution of Green AI patents.  All information was retrieved from PatentsView. In particular, we used the datasets \texttt{g\_assignee\_not\_disambiguated},  \texttt{g\_patent}, \texttt{g\_location\_disambiguated}, and \texttt{g\_cpc\_current}. Data related to AI technologies were instead extracted from the Artificial Intelligence Patent Dataset (AIPD).

To assess market concentration among patent assignees, we compute the {Gini index}, which is a standard measure of inequality. The Gini index captures the extent to which the distribution of patents deviates from perfect equality: a value of 0 indicates that all assignees hold the same number of patents, whereas a value of 1 implies that a single assignee holds all patents \cite{farris2010gini}.

Formally, let $x_i$ denote the number of patents held by assignee $i$, for $i = 1, \dots, n$, and let the values be sorted in non-decreasing order ($x_1 \leq x_2 \leq \dots \leq x_n$). The Gini index is computed as:

\begin{equation}
G = \frac{2 \sum_{i=1}^n i x_i}{n \sum_{i=1}^n x_i} - \frac{n + 1}{n}
\end{equation}

This formulation assumes a discrete distribution, as it is in our case.
Higher values of the Gini index indicate stronger concentration of innovative activity in the hands of fewer actors. We compute the index across the entire dataset and within selected time windows to track the evolution of concentration over time.

\subsection*{Topic Modeling}

In order to uncover latent themes within patent abstracts, we applied the BERTopic framework \cite{grootendorst2022bertopic}. It combines modern NLP and unsupervised learning techniques to produce interpretable topic models. BERTopic is particularly well-suited for analyzing short and domain-specific texts, such as patent abstracts, due to its modular architecture and the ability to detect an appropriate number of topics automatically.

The core steps in BERTopic include:
\begin{itemize}
    \item \textbf{Document Embeddings:} Patent abstracts were encoded using the \texttt{all-MiniLM-L6-v2} model from SentenceTransformers \cite{reimers2019sentencebert}. This model produces 384-dimensional dense vectors that capture semantic relationships between texts, offering a good trade-off between accuracy and computational efficiency. Compared to other popular models like \texttt{all-mpnet-base-v2}, \texttt{MiniLM} is approximately five times faster while preserving comparable semantic quality, and is thus widely adopted in the literature \cite{wang2021topic, kim2024naturalgas}.

    \item \textbf{Dimensionality Reduction:} The high-dimensional embeddings were projected to two-dimensional space using UMAP (Uniform Manifold Approximation and Projection) \cite{McInnes2018}. UMAP preserves the topological structure of data by minimizing the divergence between high-dimensional and low-dimensional fuzzy simplicial sets:
    \[
    \sum_{i \neq j} p_{ij} \log \frac{p_{ij}}{q_{ij}} + (1 - p_{ij}) \log \frac{1 - p_{ij}}{1 - q_{ij}},
    \]
    where \( p_{ij} \) and \( q_{ij} \) denote the pairwise affinities in the original and reduced spaces, respectively. 

\item \textbf{Clustering:} To identify groups of semantically similar abstracts, we applied HDBSCAN, a hierarchical density-based clustering algorithm \cite{8215642}, to the two-dimensional embeddings. Unlike centroid-based methods, HDBSCAN defines clusters as dense regions in the data space, allowing for the discovery of arbitrarily shaped clusters and robust handling of noise.

The algorithm constructs a hierarchy of clusters based on the mutual reachability distance:
\[
\lambda_{ij} = \frac{1}{d(x_i, x_j)},
\]
where \( d(x_i, x_j) \) is the distance between points \( x_i \) and \( x_j \) in the embedding space. From this, a so-called minimum spanning tree (MST) is created to represent the connectivity of dense regions. A MST is a graph that links all points with minimal total edge weight and no cycles \cite{8215642}.

Clusters are extracted from this hierarchy by selecting the most stable ones. The stability of a cluster \( C \) is defined as the sum of the lifetimes of its constituent points:
\[
\sum_{x \in C} (\lambda_{\text{max}}(x) - \lambda_{\text{birth}}(x)),
\]
where \( \lambda_{\text{birth}}(x) \) is the density level at which point \( x \) enters the cluster, and \( \lambda_{\text{max}}(x) \) is the level at which it leaves or the cluster dissolves. This criterion favors clusters that persist across a wide range of density thresholds.

    \item \textbf{Topic Representation:} For each cluster, BERTopic applies a class-based TF-IDF (c-TF-IDF) formulation to extract representative keywords:
    \[
    \text{c-TF-IDF}_{t,k} = \frac{f_{t,k}}{\sum_{t'} f_{t',k}} \cdot \log \left( \frac{N}{n_t} \right),
    \]
    where \( f_{t,k} \) is the frequency of term \( t \) in cluster \( k \), \( N \) is the total number of clusters, and \( n_t \) the number of clusters in which term \( t \) appears.
\end{itemize}

We remark that, prior to modeling, we performed standard pre-processing tasks: all abstracts were lowercased, tokenized, and cleaned. We removed standard English stopwords along with domain-specific boilerplate terms frequently found in patents (e.g., “herein”, “invention”) to improve the signal-to-noise ratio in topic formation. 

\paragraph{Hyperparameter Tuning and Model Selection.}
While BERTopic can infer the optimal number of topics without manual specification \cite{grootendorst2022bertopic}, the output is highly sensitive to several hyperparameters that govern UMAP and HDBSCAN. Many studies on patent corpora use the default settings (e.g., \cite{hou2025topic, kim2024naturalgas}). However, relying exclusively on default values may result in suboptimal topic separation or reduced interpretability \cite{murdock2015visualization}. A more rigorous approach involves a systematic tuning and post-hoc validation, as demonstrated in recent applications such as \cite{kim2024interdisciplinary} and \cite{kumar2024gutbrain}.

We conducted a grid search over 54 hyperparameter combinations:
\begin{itemize}
    \item \texttt{n\_neighbors} $\in \{10, 30, 50\}$ — balances UMAP’s focus between local and global structure.
    \item \texttt{min\_dist} $\in \{0.1, 0.4, 0.8\}$ — controls how tightly UMAP packs points together.
    \item \texttt{min\_cluster\_size} $\in \{50, 200, 500\}$ — HDBSCAN’s minimum cluster size.
    \item \texttt{n\_components} $\in \{2, 5\}$ — dimensionality of the UMAP projection.
\end{itemize}
We retained only models that met two criteria:
\begin{enumerate}
    \item[C1] \textbf{Outliers below 20\%:} BERTopic naturally identifies outliers, i.e., documents that are not assigned to any topic. While post-hoc reassignment strategies exist \cite{grootendorst2022bertopic}, we choose not to reassign outliers, prioritizing topic purity and interpretability, in alignment with other studies \cite{madrid2024mapping}. It has been found that high outlier rates can dilute topic clarity \cite{borcin2024optimizing}.
    \item[C2] \textbf{Topic interpretability:} We constrained the number of topics to fall between 5 and 30, as very small topic counts oversimplify the technological landscape, while very large ones lead to excessive overlap and semantic redundancy \cite{farea2024sustainable, kim2024naturalgas}.
\end{enumerate}

From the models that passed this filtering, we selected the best one based on two complementary evaluation metrics: {UMass coherence} and {topic diversity}, which together assess the internal semantic quality and distinctiveness of topics. In particular:

\begin{itemize}
    \item \textbf{UMass Coherence:} This metric, first proposed by \cite{mimno2011optimizing}, evaluates the internal consistency of each topic by measuring the extent to which its most representative terms co-occur across the documents assigned to that topic. It is computed as:
    \[
    \text{UMass}(T) = \frac{2}{n(n-1)} \sum_{i<j} \log \left( \frac{D(w_i, w_j) + \epsilon}{D(w_i)} \right),
    \]
    where \( T = \{w_1, w_2, \ldots, w_n\} \) is the set of top-$n$ terms for a topic (typically $n=10$), \( D(w_i) \) is the number of documents containing term \( w_i \), \( D(w_i, w_j) \) is the number of documents containing both \( w_i \) and \( w_j \), and \( \epsilon \) is a small constant (e.g., $10^{-12}$) to avoid division by zero. 

    UMass coherence values typically fall in the range \( [-14, 14] \), with values closer to 0 indicating higher coherence, i.e., stronger co-occurrence patterns among top words, as outlined by the official documentation of the Python library gensim \cite{oneoffcoder_topicmodeling}.
    UMass is particularly well-suited for intrinsic evaluation because it relies only on the document-term statistics of the input corpus, avoiding the need for an external reference corpus or pretrained embeddings. This makes it a natural choice in unsupervised and domain-specific settings like ours, where the vocabulary and document style differ substantially from general-purpose corpora like Wikipedia or news datasets. Other coherence metrics—such as C\_V, NPMI, or C\_UCI—require word co-occurrence statistics from a large external corpus or the use of context-based similarity scores (e.g., cosine similarity between word vectors), which may introduce bias or misalignment when applied to highly technical texts like patents \cite{ko2021patents, zain2024bert}.
    Furthermore, UMass coherence is tightly aligned with the actual topic assignment procedure used by BERTopic, which clusters documents and computes topic-term importance using c-TF-IDF scores over the same corpus. This consistency strengthens the interpretability and methodological robustness of the model evaluation.

    \item \textbf{Topic Diversity:} This metric assesses the distinctiveness of the topics by measuring the proportion of unique terms among the top-ranked words across all topics. It is defined as:
    \[
    \text{Diversity} = \frac{|\bigcup_{k=1}^{K} W_k|}{K \cdot n} \times 100\%,
    \]
    where \( K \) is the total number of topics, \( W_k \) is the set of top-$n$ words for topic $k$ (typically $n=10$), and \( |\bigcup_{k=1}^{K} W_k| \) is the total number of unique words across all topic descriptors. A higher topic diversity indicates that the model has discovered a wider variety of topics with less lexical overlap. In contrast, low diversity may signal redundancy, in the sense of topics that are lexically similar and semantically overlapping, resulting in poor differentiation and reduced utility for interpretation or policy insights.

\end{itemize}

By jointly optimizing for high UMass coherence and high topic diversity, we aim at our selected model to generate topics that are both semantically well-formed and clearly distinct from one another. 
Furthermore, running BERTopic multiple times, even with the same parameters, leads to slightly different results in terms of the number of topics, outliers and documents per topic. This is an important point to keep in mind, stemming from the stochastic nature of UMAP. Following \cite{kumar2024gutbrain}, we conducted consistency checks and observed high stability in the identified topic structures and their technological coherence.

The selected  model had \texttt{n\_neighbors=30}, \texttt{min\_dist=0.1}, \texttt{min\_cluster\_size=200}, and \texttt{n\_components=2}, producing 16 topics, \np{7656} outliers (12.1\%), a UMass coherence of $-2.9693$, and topic diversity of 81.25\%.

\subsection*{UMAP-based 2D representation, and analysis of the temporal evolution of technological domains}

Following topic extraction via BERTopic, we performed a qualitative grouping of the resulting topics (or technological domains in our interpretation) into four broader macro-domains. This was based on the spatial distribution of topics in the two-dimensional UMAP projection, which preserves local semantic relationships between documents, allowing us to visually cluster them into thematically coherent regions \cite{McInnes2018}. Each macro-domain corresponds to a distinct area of innovation within the Green AI landscape.

To investigate temporal trends in technological activity, we computed the number of patents associated with each domain on a yearly basis. Given annual fluctuations and data sparsity in some periods, we applied a rolling average with a 3-year window to smooth short-term variations and highlight structural trends. For each year \( t \), the rolling average of patent counts \( \bar{y}_t \) is defined as:
\[
\bar{y}_t = \frac{1}{3} \sum_{i = t-1}^{t+1} y_i,
\]
where \( y_i \) denotes the number of patents assigned to the macro-domain in year \( i \). This smoothed time series enables a clearer visualization of growth patterns and inflection points in each macro-domain over time.

\subsection*{Impact analysis with forward citations and stock market value}
To evaluate the technological and economic relevance of patents, we carried out a twofold analysis. 
First, we used the number of \texttt{g\_us\_patent\_citation} from PatentsView as a proxy for technological impact. This variable captures the number of forward citations received by a patent from subsequent U.S. patents, a standard indicator of its influence on later technological developments \cite{AristodemouTietze2018}.
Second, we assessed the economic value of patents by integrating the firm-level patent valuation model proposed by \cite{KoganPapanikolaouSeruStoffman2017}. This model estimates the market value of a patent \( \xi_j \) as a function of the firm's abnormal stock return around the patent grant date. Specifically, for each patent \( j \), the value is computed as:

\[
\xi_j = \frac{1}{1 - \bar{\pi}} \cdot \frac{1}{N_j} \cdot \mathbb{E}[v_j \mid R_j] \cdot M_j,
\]
where \( \bar{\pi} \approx 0.56 \) is the average grant rate, \( N_j \) is the number of patents granted to the same firm on the same day, \( M_j \) is the firm's market capitalization on the day before grant, and \( \mathbb{E}[v_j \mid R_j] \) is the filtered component of the idiosyncratic return \( R_j \) attributable to the patent. The expectation is derived under the assumption that the true patent value \( v_j \) follows a truncated normal distribution and that noise \( \varepsilon_j \) is normally distributed:
%
$R_j = v_j + \varepsilon_j$.

To isolate the signal, the model introduces a signal-to-noise ratio \( \delta \), calibrated from event-study regressions on return volatility. The filtered expectation is then given by:

\[
\mathbb{E}[v_j \mid R_j] = \delta R_j + \sqrt{\delta} \sigma_{\varepsilon} \cdot \frac{\phi\left(-\sqrt{\delta} \cdot \frac{R_j}{\sigma_{\varepsilon}} \right)}{1 - \Phi\left(-\sqrt{\delta} \cdot \frac{R_j}{\sigma_{\varepsilon}} \right)},
\]
where \( \phi \) and \( \Phi \) denote the standard normal PDF and CDF, respectively. In our analysis, we use precomputed values of \( \xi_j \) made available by the authors, which apply this methodology to link patent events to firm-level stock returns and correct for noise and grant probability biases.

This combined approach enabled us to quantify both the technological diffusion and economic significance of Green AI patents.

\paragraph{Data availability.}
 The code and dataset supporting this study have been deposited in Zenodo at \url{https://doi.org/10.5281/zenodo.15545360} (more info in ref \cite{Emer2025GreenAIData}). 

\bibliography{sample}



\section*{Acknowledgements}

The work has been partially supported by project SMaRT COnSTRUCT (CUP J53C24001460006), in the context of FAIR (PE0000013, CUP B53C22003630006) under  the National Recovery and Resilience Plan (Mission 4, Component 2, Line of Investment 1.3) funded by the European Union - NextGenerationEU.

\section*{Author contributions statement}

L.E. contributed to all aspects of the research, including conceptualization, methodology, writing, and conducted the computational analyses. A.M. and A.V. contributed equally to research conceptualization, methodology, and writing. All authors have read and approved the final version of the manuscript.

\section*{Additional information}
\noindent
\textbf{Competing interests} The authors declare no competing interests.  


\end{document}